\pdfoutput=1 
\documentclass[aps,prb,twocolumn,superscriptaddress,showpacs]{revtex4-1} 
\usepackage{amsmath,amssymb,mathrsfs}
\usepackage{latexsym}
\usepackage{graphicx} 
\usepackage{epstopdf}
\usepackage{graphicx,epstopdf,color}
\usepackage{amsfonts}
\usepackage{amsmath,amssymb,mathrsfs}
\usepackage{bm}

\newcommand{\bs}[1]{\boldsymbol{#1}}

\newcommand{\up}{\uparrow}
\newcommand{\dw}{\downarrow}

\newcommand{\pd}{{\phantom{\dag}}}

\def\eg{\emph{e.g.}\ }
\def\ea{\emph{et al.\ }}

\allowdisplaybreaks[1]
\begin{document}
\title{Magnetic ordering phenomena of interacting quantum spin Hall models}

\author{Johannes Reuther}
\affiliation{Department of Physics and Astronomy, University of California, Irvine, CA 92697, USA}
\author{Ronny Thomale}
\affiliation{Department of Physics, Stanford University, Stanford, CA 94305, USA}
\author{Stephan Rachel}
\affiliation{Department of Physics, Yale University, New Haven, CT 06520, USA}
\affiliation{Institute for Theoretical Physics, Dresden University of Technology, 01062 Dresden, Germany}
 \pagestyle{plain}

\begin{abstract}
The two-dimensional Hubbard model defined for topological band structures exhibiting a quantum spin Hall effect poses fundamental challenges in terms of phenomenological characterization and microscopic classification. In the limit of infinite coupling $U$ at half filling, the spin model Hamiltonians resulting from a strong coupling expansion show various forms of magnetic ordering phenomena depending on the underlying spin-orbit coupling terms. We investigate the infinite $U$ limit of the Kane--Mele Hubbard model with $z$-axis intrinsic spin-orbit coupling as well as its generalization to a generically multi-directional spin orbit term which has been claimed to account for the physical scenario in monolayer $\text{Na}_2\text{IrO}_3$. We find that the axial spin symmetry which is kept in the former but broken in the latter has a fundamental impact on the magnetic phase diagram as we vary the spin orbit coupling strength. While the Kane--Mele spin model shows a continuous evolution from conventional honeycomb N\'eel to $XY$ antiferromagnetism which avoids the frustration imposed by the increased spin-orbit coupling, the multi-directional spin-orbit term induces a commensurate to incommensurate transition at intermediate coupling strength, and yields a complex spiral state with a 72 site unit cell in the limit of infinite spin-orbit coupling. From our findings, we conjecture that in the case of broken axial spin symmetry there is a large propensity for an additional phase at sufficiently large spin-orbit coupling and intermediate $U$. 
\end{abstract}

\pacs{31.15.V-, 75.10.Jm, 03.65.Vf}    

\maketitle

\section{Introduction}

The discovery of the quantum Hall effect has initialized the era of topological phases in condensed matter physics. For non-interacting band structures with topologically unconventional properties, topological indices take over the role of conventional order parameters and can be linked to quantization phenomena of edge modes measured in experiment. The first example of such an index was introduced by Thouless, Kohmoto, Nightingale, and den Nijs (TKNN) for the integer quantum Hall effect (IQHE).\cite{thouless-82prl405} They could show that the first Chern number---the TKNN invariant---is proportional to the transversal Hall conductivity $\sigma_{xy}$ which is  the integral of the Berry curvature over the Brillouin zone. Nearly a decade ago after Haldane realized that one can define lattice versions of IQHE called Chern insulators where complex hopping breaks time-reversal symmetry\cite{haldane88prl2015}, the most recent example of a non-interacting topological state of matter is the topological insulator\cite{hasan-10rmp3045,qi-11rmp1057,moore10n194}. It is characterized by a $\mathbb{Z}_2$ topological index\cite{kane-05prl146802,kane-05prl226801}.
$\mathbb{Z}_2$ topological insulators (TIs) have not only been proposed theoretically\cite{kane-05prl146802,kane-05prl226801,bernevig-06s1757} but have also been found in subsequent experiments.\cite{koenig-07s766} 
The minimal model of a $\mathbb{Z}_2$ topological insulator is a four--band model possessing a finite $\mathbb{Z}_2$ invariant, which in its simplest form is a minimal time--reversal invariant generalization of a Chern insulator. All two--dimensional band structures exhibiting a non-trivial $\mathbb{Z}_2$ invariant can be adiabatically transformed into each other, i.e. without closing the bulk gap. In contrast, transforming a $\mathbb{Z}_2$ TI phase into any other topologically trivial phase causes a quantum phase transition where the bulk gap must close. To date, these topological band insulators are well understood and systematically classified by symmetry.\cite{schnyder-08prb195125,kitaev09aipcp22} 

As soon as interactions are taken into account, the full scope of possible scenarios extends to (i) topological band structure phases where the interactions would only renormalize the band parameters but do not change the topology along with (ii) conventional ordering phenomena where all features of the topologically non-trivial phase are gone\,\cite{pesin-10np376,rachel-10prb075106,kargarian-11prb165112,hohenadler-11prl100403,zheng-11prb205121,hohenadler-12prb115132,vaezi-12prb195126,budich-arxiv1203,lee-11prl166806,griset-12prb045123,wang-10prl256803,gurarie11prb085426,krempa-10prb165122}, and (iii) topological Mott insulators,\,\cite{pesin-10np376,rachel-10prb075106,kargarian-11prb165112,raghu-08prl156401} and
 (iv) topological bulk order driven by strong interactions along with finite quantum dimension~\cite{wen-89prb7387}, fractionalization of quantum numbers~\cite{su-79prl1698,laughlin83prl1395}, and fractional statistics~\cite{leinaas-77ncb1} as well as interaction-driven topological band structure phases which are not adiabatically connected to the non-interacting limit~\cite{raghu-08prl156401,kargarian-11prb165112,budich-arxiv1203}. In analogy to the non-interacting counterpart, topological bulk order was first discovered in the fractional quantum Hall effect (FQHE)~\cite{tsui-82prl1559,laughlin83prl1395} before the concept of topological order was established by Wen~\cite{wen-89prb7387}.
Due to the diversity of possible phases even in the same symmetry sector, a general classification for interacting topological phases is lacking so far. 
Aside from many other challenges, it is of particular interest whether the concept of a topological band structure and topological bulk order can both manifest itself in a single microscopic model~\cite{fiete12pe845}. For example, competing magnetic fluctuations originating from a  topological band structure model could manage to stabilize a topological spin liquid phase. This is the general motif of a class of scenarios which we further investigate in this article.

\begin{figure}[t]
\centering
\includegraphics[scale=0.62]{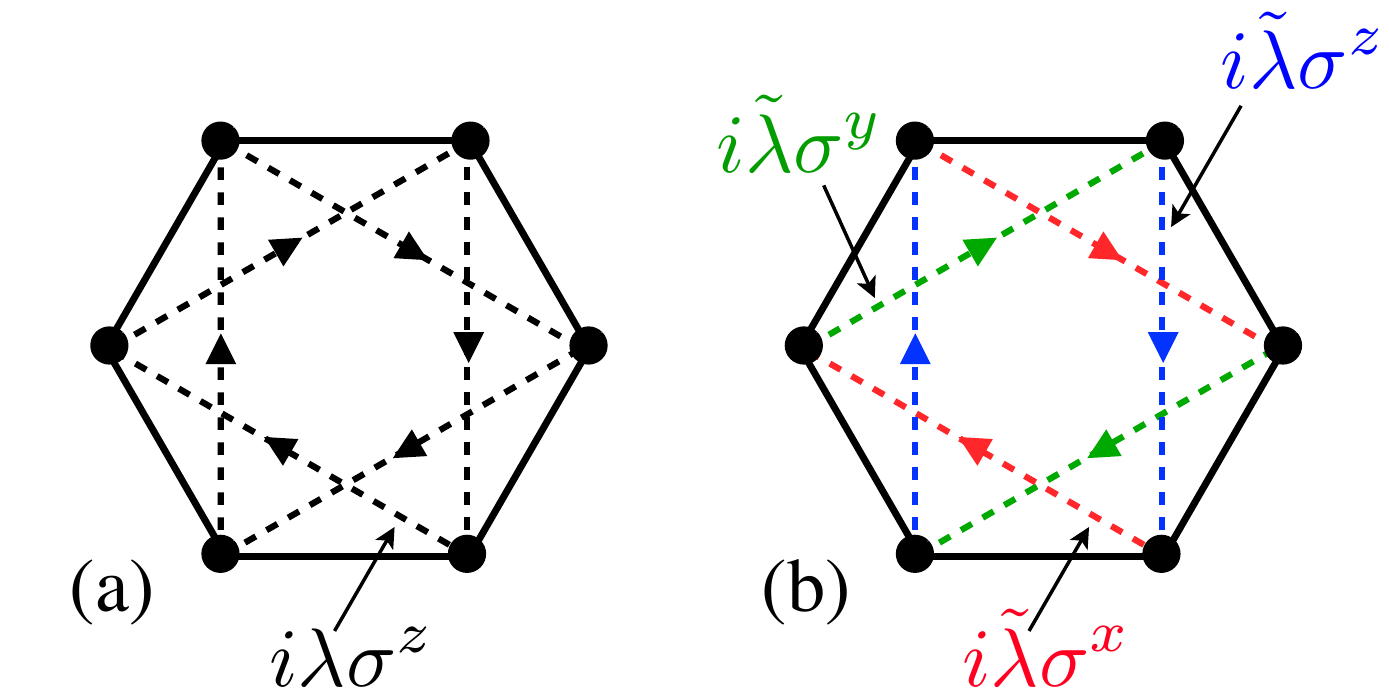}
\caption{Intrinsic spin orbit terms with amplitude $\lambda$ according (a) to \eqref{ham:km} for the Kane--Mele model  and (b) to \eqref{ham:shitade} for the SI model with multi-directional SOC amplitude $\tilde{\lambda}$.}
\label{fig:soc-hoppings}
\end{figure}

As we will show in detail in the following, the presence or absence of axial spin symmetry stemming from the topological band structure in the interacting case will crucially determine the magnetic order and disorder phenomena which appear in the strong coupling limit. Generically, the full $\text{SU}(2)$ is broken for interacting topological band structure models because of spin orbit coupling terms. Still, it is both possible that the spin orbit terms break $\text{SU}(2)$ down to $\text{U}(1)$, leaving a continuous axial spin symmetry intact, or completely break spin rotation symmetry. 
Since its custodial time-reversal symmetry is unaffected, it is irrelevant for the $\mathbb{Z}_2$ index of the weakly coupled model whether the axial spin symmetry of the TI is conserved or not: although it has been shown recently that breaking of axial spin symmetry causes a momentum--dependent rotation of the spin quantization axis of the helical edge states,\cite{schmidt-12prl156402} the topological band structure with  conserved spin symmetry can still be transformed into one with broken spin symmetry without closing of the bulk gap. In contrast, for strong interactions, the resulting phase diagram crucially depends on presence or absence of axial spin symmetry; more specifically, it was claimed that the combination of strong interactions and strong spin orbit coupling might give rise to a topologically ordered phase on the honeycomb lattice when spin is not conserved. This would then be a paradigmatic candidate model which includes both a topological band structure phase and topological bulk order in its phase diagram.\cite{ruegg-12prl046401}  
Unfortunately, only the conserved U(1) symmetry appears to open up the possibility to successfully perform quantum Monte Carlo (QMC) simulations for the regime of intermediately coupled topological band structure models; when this symmetry is absent, we instead have to rely on limited mean--field, slave--particle, or other approximate methods. 

In our work, we propose the strategy to first gain insight about this kind of models in the limit of infinitely large interactions on the footing of an accurate method adapted to this limit, and to find out which of the approximate results at intermediate interaction strength is compatible with it. For this purpose, we employ pseudofermion functional renormalization group (PFFRG) which has been recently developed and employed by two of us in the context of various models of frustrated magnetism~\cite{reuther-10prb144410,reuther-11prb024402,reuther-11prb014417,reuther-11prb100406,singh-12prl127203}. In particular, the anisotropic spin terms do not pose additional challenges to the performance of the PFFRG, which at the same time allows us to study large system sizes beyond any other microscopic numerical procedure for two-dimensional spin models.  

In this paper, we investigate the strong coupling limit of two different topological band structures accompanied with Hubbard onsite interactions on the honeycomb lattice: the Kane--Mele (KM) model\cite{kane-05prl146802,kane-05prl226801} preserving axial spin symmetry and a related model which was proposed in the context of Na$_2$IrO$_3$ by Shitade \ea\cite{shitade-09prl256403} which explicitly breaks axial spin symmetry. Because of its connection to sodium iridate, it will be referred to as SI model in the following. We find that while magnetism in the presence of axial spin symmetry can generically avoid the frustration effects caused by the anisotropic spin terms induced by spin-orbit coupling and generically yields commensurate magnetism, the broken axial spin symmetry scenario naturally leads to commensurate-incommensurate transitions and, as a consequence, a much more complex magnetic phase diagram. As such, we conjecture that the latter scenario will be most promising to stabilize unconventional, possibly topologically bulk ordered phases resulting from anisotropic spin terms.
We also discuss our findings in the context of recent results\cite{ruegg-12prl046401} for the corresponding Hubbard models at finite coupling.

The paper is organized as follows. In Section \ref{sec:topo-bands}, we introduce the KM and SI models and discuss their main properties. The mean field phase diagrams of the corresponding Hubbard models -- the Kane--Mele--Hubbard (KMH) model as well as the sodium iridate Hubbard (SIH) model -- are briefly reviewed in Section \ref{sec:correlated-ti}. We subsequently introduce the corresponding spin models in Section \ref{sec:scl}. In Section \ref{sec:method}, we elaborate on the PFFRG method which we employ to investigate the magnetic phase diagrams of the KM and SI spin models the results of which are presented in Section \ref{sec:results}.
In Section \ref{sec:discussion}, we draw a line from our findings at infinite coupling to the corresponding Hubbard models at finite coupling  in the context of the recently proposed QSH$^\star$ phase, a topologically ordered phase in the SIH model.\cite{ruegg-12prl046401} In particular, we also point out important generalizations of our study with respect to Rashba coupling, which will generically break axial spin symmetry. 
In Section \ref{sec:conclusion}, we conclude that the role of the axial spin symmetry is crucial to characterize magnetic order and disorder phenomena of interacting topological honeycomb band structures and leads to a better understanding of the general theme of interaction effects in topological insulators.

Throughout this paper we use the following notations: the non--interacting topological insulators, i.e. the band structures are denoted by $h_{\rm KM}$ and $h_{\rm SI}$, respectively. The corresponding Hubbard models are called $H_{\rm KM}$ and $H_{\rm SI}$ while the spin models are denoted by $\mathcal{H}_{\rm KM}$ and $\mathcal{H}_{\rm SI}$, respectively. The real nearest neighbor hopping amplitude is $t$; the intrinsic spin orbit couplings are called $\lambda$ for the KM model and $\tilde\lambda$ for the SI model.

\section{Topological band structures}\label{sec:topo-bands}

The QSH honeycomb models are particularly accessible from a theoretical perspective: as there are already two sites per unit cell, it is sufficient to study a single orbital scenario where complex hoppings generate the band inversion giving rise to a non-trivial $\mathbb{Z}_2$ invariant. There is hope that the QSH effect on the honeycomb lattice might be realized, \eg by doping heavy adatoms in graphene\cite{weeks-11prx021001} or by using silicene\cite{liu-11prl076802} which has recently been accomplished experimentally\cite{vogt-12prl155501}. Depending on the concise form of the spin-orbit coupling terms, the axial spin symmetry may or may not be broken in the interacting case. In this section we briefly introduce the two representative models for both scenarios which are subject to further investigation in the following.

\subsection{Kane--Mele model}
Kane and Mele\,\cite{kane-05prl146802,kane-05prl226801} proposed the quantum spin Hall (QSH) effect in graphene based on symmetry consideration. They realized that a mass term $\propto \sigma^z\tau^z\eta^z$ does not violate any symmetries of graphene and thus must be allowed. Here, $\sigma$ is associated with the electron spin, $\tau$ with the valleys, and $\eta$ with the sublattices. 
The Kane--Mele model is governed by the tight--binding Hamiltonian
\begin{equation}
\label{ham:km}\begin{split}
h_{\rm KM}=-t \sum_{\langle ij\rangle\sigma} c_{i\sigma}^\dag c_{j\sigma}^\pd
+i\lambda \sum_{\ll ij \gg}\sum_{\alpha\beta} \nu_{ij} c_{i\alpha}^\dag \sigma^z_{\alpha\beta} c_{j\beta}^\pd\ 
\end{split}
\end{equation}
In principle, there is also the Semenoff mass term which we will ignore for the moment. 
Similarly, the Rashba spin orbit term with amplitude $\lambda_{\rm R}$ is neglected unless noted otherwise.
The first term in~\eqref{ham:km} is the usual nearest--neighbor hopping on the honeycomb lattice giving rise to the Dirac band structure. The second term in~\eqref{ham:km} is the lattice version of the $\sigma^z\tau^z\eta^z$--term (a second neighbor hopping) which corresponds to an intrinsic spin orbit coupling (SOC). The convention of this hopping is illustrated in Fig.\,\ref{fig:soc-hoppings}a. 
The nearest neighbor hopping term preserves the $C_{6v}$ lattice symmetry of the honeycomb lattice as well as SU(2) symmetry of the electron spin. The intrinsic SOC reduces the lattice symmetry to $C_{3v}$ and the spin symmetry to U(1). 
Any finite $\lambda$ opens the gap of the Dirac band structure and gives rise to QSH effect, i.e.  to a topological insulator phase characterized by a finite $\mathbb{Z}_2$ invariant, or, in this case, Chern number for each spin species. This situation is very special since the Hamiltonian fully decouples into two independent Chern insulators with opposite Hall conductivity. Generically, we expect the presence of additional terms breaking the U(1) spin symmetry and mixing the spin channels. The Rashba term is such an additional term which will be further commented on in Section~\ref{sec:discussion}. Even for finite Rashba coupling $\lambda_R$, however, the QSH phase is stable as long as $\lambda_R < 2\sqrt{3}\lambda$.\cite{kane-05prl146802}

\begin{figure*}[t]
\centering
\includegraphics[scale=0.7]{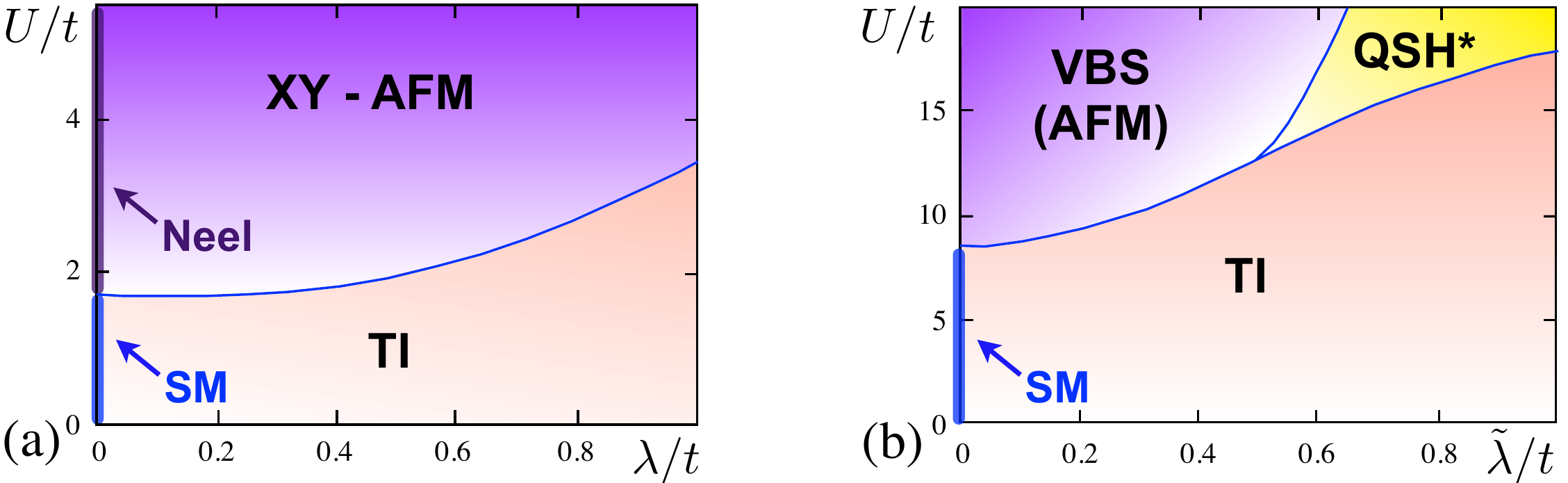}
\caption{(color online). (a) phase diagram of the Kane--Mele--Hubbard model as obtained in 
Ref.\,\onlinecite{rachel-10prb075106}. The transition from a topological insulator (TI) to $XY$-plane antiferromagnet (AFM) was derived within slave-rotor theory which underestimates $U_c$. (b) mean field phase diagram of the sodium--iridate Hubbard model as obtained in Ref.\,\onlinecite{ruegg-12prl046401}. The transition from TI to a valence bond solid (VBS) phase that links to AFM was derived within slave-spin theory which overestimates $U_c$. The phase diagram of (b) is qualitatively similar to (a) apart from the additional ``QSH$^\star$ phase". At $\lambda=\tilde\lambda=0$ and not too large $U$ the semi-metal (SM) phase of graphene is present. 
See main text for details.}
\label{fig:kmh+sih}
\end{figure*}

\subsection{Sodium iridate tight binding model}
Soon after Kane and Mele's milestone works, it turned out that the spin orbit gap in graphene is vanishingly small. Therefore other materials with effective honeycomb structure were considered
as candidates for the QSH effect as proposed by Kane and Mele. In 2008, Shitade \ea\cite{shitade-09prl256403} came up with the sodium iridate Na$_2$IrO$_3$  as a layered honeycomb system. The authors claimed that the QSH effect might be realized if Coulomb interactions are not too strong. A monolayer was shown to be described by a Kane--Mele-type Hamiltonian. The intrinsic spin orbit coupling was assumed to be relatively large due to the heavier iridium atoms in contrast to graphene's carbon atoms. Assuming trivial hybridization between nearest neighbor Ir atoms, Shitade \ea found an intrinsic SOC being similar but different to the KM SOC. It depends on the direction of the spin orbit hopping whether the spin degree of freedom is associated with $\sigma^x$, $\sigma^y$, or $\sigma^z$.
The sodium iridate model is governed by the Hamiltonian
\begin{equation}\label{ham:shitade}
h_{\rm SI}=-t \sum_{\langle ij\rangle\sigma} c_{i\sigma}^\dag c_{j\sigma}^\pd + i\tilde\lambda
\sum_{\ll ij \gg_\gamma}\sum_{\alpha\beta} c_{i\alpha}^\dag \sigma^\gamma_{\alpha\beta}  c_{j\beta}^\pd\ , 
\end{equation}
where $\gamma=x,y,z$ is associated with the different next--nearest neighbor links on the honeycomb lattice (Fig.\,\ref{fig:soc-hoppings}b). The main difference of this generalized SOC compared to the KM SOC is that axial spin symmetry is not conserved. 
As for the KM model, infinitesimally small $\tilde\lambda$ opens the gap at the Dirac cones and causes QSH effect.

The band structures of $h_{\rm KM}$ and $h_{\rm SI}$  both belong to the $\mathbb{Z}_2$ universality class and are thus adiabatically connected.
Both systems exhibit helical edge states on open geometries such as cylinder or disk.

\section{Correlated topological insulators}\label{sec:correlated-ti}

Let us now add Hubbard onsite interactions, 
\begin{equation}\label{ham:int}
H_I=U\sum_i n_{i\up}n_{i\dw}
\end{equation}
which yields rich phase diagrams for both band structures. 
While the $U$--$\lambda$ phase diagram of the KMH model is well understood\,\cite{rachel-10prb075106,hohenadler-11prl100403,zheng-11prb205121,yu-11prl010401,wu-12prb205102,budich-arxiv1203}, the $U$--$\tilde\lambda$ phase diagram of the SIH model is rarely studied\cite{shitade-09prl256403,ruegg-12prl046401}, and the available results are controversial. In the following, we will briefly review the phase diagrams of both Hubbard-type models.

\subsection{Kane--Mele--Hubbard model}

The KMH model is described by a combination of the KM and Hubbard model,
\begin{equation}
\label{kmhubbard}
H_{\rm KM}= h_{\rm KM} + H_I\ .
\end{equation}
In Ref.\,\onlinecite{rachel-10prb075106} the phase diagram shown in Fig.\,\ref{fig:kmh+sih}a was derived through slave rotor theory. The semi--metal (SM) phase of graphene ($\lambda=0$) as well as the topological insulator phase ($\lambda\not= 0$) are stable up to moderate interactions. Above a critical interaction strength $U_c$, one finds an antiferromagnetically ordered phase which is of N\'eel type ($\lambda=0$) or of $XY$--type ($\lambda\not= 0$), respectively. 
At $\lambda=0$ and intermediate $U$, a quantum spin liquid phase has been proposed\,\cite{meng-10n847} recently; this conjecture has been challenged lately.\,\cite{sorella-12arXiv:1207.1783} For very small $\lambda$ it survives but eventually vanishes for $\lambda\leq 0.05 t$\cite{hohenadler-11prl100403,zheng-11prb205121,yu-11prl010401,wu-12prb205102}.
 Since the spin liquid is destroyed by finite $\lambda$ and just a remnant of the non--topological $\lambda=0$ case, we omit the phase here for clarity.
Also for the strong coupling analysis in this paper we will assume that we are deep in the strong coupling regime where this intermediate coupling phenomenon is irrelevant for our analysis.

\subsection{Sodium iridate Hubbard model}
Recently, R\"uegg and Fiete have studied the SIH model\,\cite{ruegg-12prl046401} governed by the Hamiltonian
\begin{equation}
H_{\rm SI} = h_{\rm SI} + H_I\ .
\end{equation}
They used a $\mathbb{Z}_2$ slave--spin mean--field approach and proposed an interesting phase diagram (Fig.\,\ref{fig:kmh+sih}b). It is similar to the KMH model, while there is an additional phase for large SOC $\tilde\lambda$ and large $U$, dubbed QSH$^\star$ phase, which presumably extends to the strong coupling regime. Note that this is not a quantum spin Hall phase, but a topological liquid which is characterized by a four--fold degeneracy on a torus, where the elementary excitations are fractional particles obeying Abelian statistics.
Recently it was questioned, however, whether the employed $\mathbb{Z}_2$ slave spin approach is justified.\cite{nandkishore-12prb045128}
Also, within the $\mathbb{Z}_2$ slaveÐspin approach one cannot find local moments such as an antiferromagnetically ordered phase (AFM), but instead obtains a valence bond solid (VBS) phase.  In the limit $\tilde\lambda\to 0$ it is obvious that one should find Neel order instead and that the VBS order is an artifact of the specific slave particle approach.

Regarding the values of $U_c$ (\eg for $\lambda=\tilde\lambda=0$), one should keep in mind that the {\it microscopic} $U_c \sim 4.3$ as found within QMC\,\cite{meng-10n847} is understimated by slave rotor theory ($U_c = 1.68$) while it is overestimated by the slave spin approach ($U_c \sim 8$) (Fig.\,\ref{fig:kmh+sih}).

\section{Strong coupling limit}\label{sec:scl}
We consider the limit of infinitely strong electron--electron interactions. As a result, charge fluctuations are frozen out and we obtain a pure spin Hamiltonian at half filling. Most importantly, the complex next--nearest neighbor spin orbit hoppings result in anisotropic and more complicated second neighbor spin exchange terms which we analyze in the following.

\subsection{Kane--Mele spin model}
Taking the limit $U\to\infty$ of the Kane--Mele--Hubbard model~\eqref{kmhubbard} results in the effective spin model\,\cite{rachel-10prb075106} 
\begin{equation}\label{kanemelespinmodel}
\mathcal{H}_{\rm KM}=
J_1 \sum_{\langle ij \rangle} \bs{S}_i \bs{S}_j + J_\lambda \!\!\sum_{\ll ij \gg}\!\left[ -S_i^x S_j^x - S_i^y S_j^y + S_i^z S_j^z \right]
\end{equation}
where $J_1=4t^2/U$ and $J_\lambda=4\lambda^2/U$. The second neighbor exchange term (indicated by $\ll\cdot\gg$) acting merely on individual, i.e. triangular sublattices partially frustrates the system. The $XY$-spin terms prefer ferromagnetic order on the individual sublattices which is consistent with antiferromagnetic order on the original honeycomb lattice; in contrast, the 
Ising term $S_i^zS_j^z$ favors antiferromagnetic order on the sublattice competing with both the $XY$-terms and the $J_1$ term. The magnetization, which might point in any direction for $J_{\lambda}=0$ due to spin rotational invariance, turns into the $XY$-plane in order to avoid the frustrating part of the $J_{\lambda}$ term\,\cite{rachel-10prb075106}. These findings were confirmed within QMC\,\cite{hohenadler-11prl100403,zheng-11prb205121}, 
variational cluster approximation (VCA)\,\cite{yu-11prl010401}, and cluster dynamical mean--field theory (CDMFT)\,\cite{wu-12prb205102} calculations at intermediate $U/t\approx 5\ldots 9$ and small $\lambda$. For small $J_\lambda$, one can thus employ $\mathcal{H}_{\rm KM}$ to compare other numerical approaches against PFFRG method, which we will use in the following.

\subsection{Sodium iridate spin model}
The strong coupling limit of the SIH model is given by the spin Hamiltonian 
\begin{equation}\label{ham:qshstar-spin}
\mathcal{H}_{\rm SI}=J_1\sum_{\langle ij \rangle} \bs{S}_i \bs{S}_j - J_{\tilde\lambda} \sum_{\ll ij \gg} \bs{S}_i\bs{S}_j
+2J_{\tilde\lambda} \!\! \sum_{\gamma-{\rm links}} \!\! S_i^\gamma S_j^\gamma\ .
\end{equation}
Note that the $\gamma$--links  are the second neighbor links (the green, red, and blue lines in Fig.\,\ref{fig:soc-hoppings}b).
It is structurally similar to the Heisenberg--Kitaev (HK) Hamiltonian\;\cite{chaloupka-10prl027204,reuther-11prb100406} which has been found to adequately describe the A$_{2}$IrO$_{3}$ iridates from a spin-orbit Mott picture (A=Na or Li)~\cite{singh-12prl127203}. Whereas the SI model assumes the nearest-neighbor hybridization to be trivial and to be essentially given by real Ir-Ir hybridization, the kinetic theory underlying the HK model more carefully resolves the emergent terms from a multi-orbital Ir-O cluster superexchange model~\cite{jackeli-09prl017205,chaloupka-10prl027204}. Depending on the Ir-O-Ir angle, these terms are either more or less relevant than the next nearest neighbor exchange terms which are considered in the SI model~\cite{reuther-11prb100406}. For the links in vertical direction (links with $\sigma^z$), one obtains the same term as for $\mathcal{H}_{\rm KM}$, while for the links associated with $\sigma^x$ one finds 
$+S_i^x S_j^x - S_i^y S_j^y - S_i^z S_j^z$ and so on. For $\mathcal{H}_{\rm KM}$ we have seen that the magnetization turns into the $XY$-plane. Here, however, the term $+S_i^x S_j^x - S_i^y S_j^y - S_i^z S_j^z$ will force the magnetization into the $YZ$-plane while the term $-S_i^x S_j^x + S_i^y S_j^y - S_i^z S_j^z$
favors the $XZ$-plane and so on. Since all the terms (links) are equally distributed over the lattice, a priori no plane or direction is preferred. As the system sets out to be N\'eel--ordered for $J_{\tilde{\lambda}}=0$, it is conceivable that the competing ordering tendencies might at first compensate each other and allow for a persistent N\'eel order at small  $J_{\tilde{\lambda}}$.

\section{Method}\label{sec:method}

The PFFRG approach~\cite{reuther-10prb144410,reuther-11prb024402,reuther-11prb014417,reuther-11prb100406,gottel-12prb214406} starts by reformulating the spin Hamiltonian in terms of a pseudo fermion representation of the spin-1/2 operators $S^{\mu} = 1/2 \sum_{\alpha\beta} f_{\alpha}^{\dagger} \sigma_{\alpha\beta}^{\mu} f_{\beta}$, ($\alpha,\beta = \uparrow,\downarrow$, $\mu = x,y,z$) with fermionic operators $f_{\uparrow}$ and $f_{\downarrow}$ and Pauli-matrices $\sigma^{\mu}$. Such a representation enables us to apply Wick's theorem, leading to standard Feynman many-body techniques. In this pseudofermion language, quantum spin models become strongly coupled models with zero fermionic bandwidth and finite interaction strength.

A major advancement of the PFFRG~\cite{reuther-11prb024402} is that it allows to tackle this situation by providing a systematic scheme for the infinite order self-consistent resummations. The first conceptual step is the introduction of an  infrared frequency cutoff $\Lambda$ in the fermionic propagator. The FRG then formulates differential equations for the evolution of all $m$-particle vertex functions under the flow of $\Lambda$~\cite{metzner-12rmp299}. Hence, one might think of the diagrammatic summations as being performed during the RG flow: each discretized RG step effectively increases the amount of diagrams included in the approximation. 

To reduce the infinite hierarchy of coupled equations to a closed set, a common approach is to restrict oneself to one-loop diagrams. The PFFRG extends this approach by also including certain two-loop contributions~\cite{katanin04prb115109,reuther-11prb024402} to retain a sufficient backfeeding of self-energy corrections to the two-particle vertex evolution. A crucial property of the PFFRG is that the the two-particle vertex includes both graphs that favor magnetic order and those that favor disorder in such a way that the method treats both tendencies on equal footing~\cite{reuther-11prb024402}. It is the two-particle vertex which allows to extract magnetic susceptibility as the main outcome of the PFFRG.
\begin{figure}[t]
\centering
\includegraphics[scale=0.6]{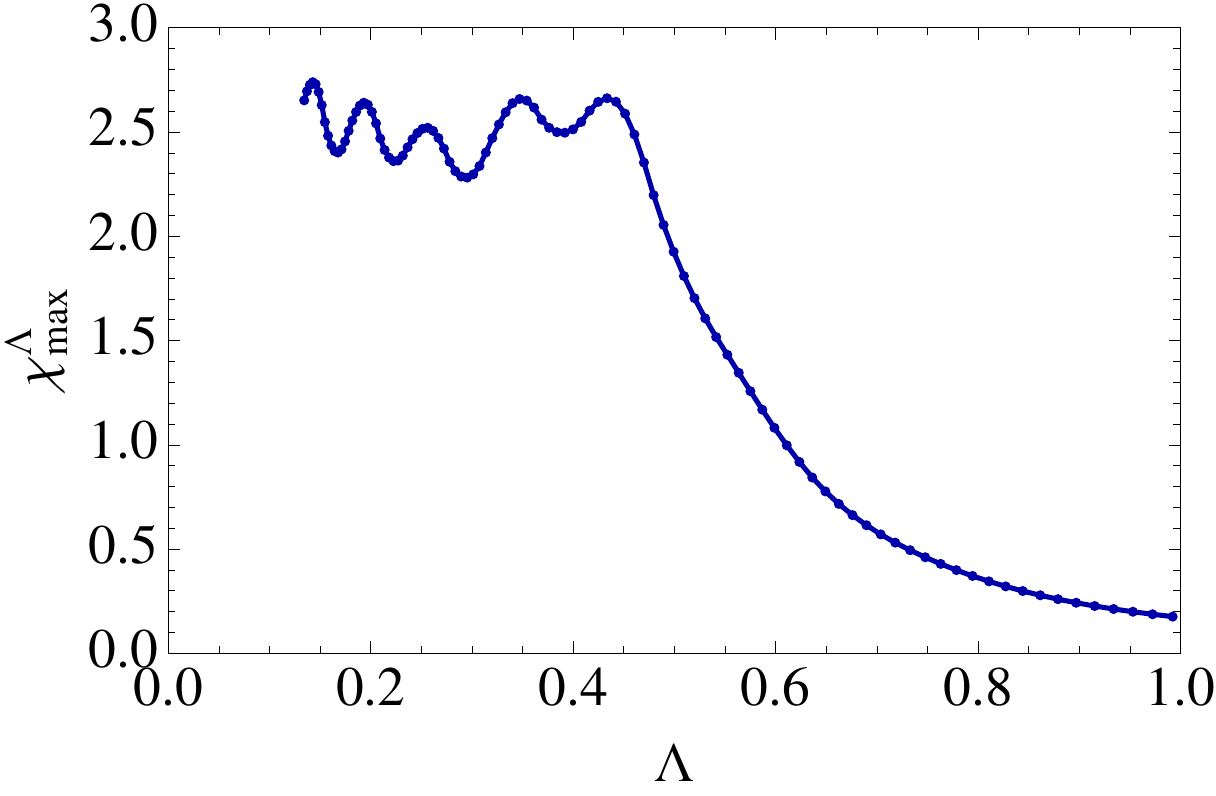}
\caption{Characteristic behavior of the flowing ($\Lambda$-dependent) susceptibility in a magnetically ordered phase. While the RG flow is smooth above some critical value $\Lambda_{\text{c}}\approx0.45$, a numerically unstable regime is found below that value. This feature signals a magnetic instability which becomes a divergence in the thermodynamic limit. The specific case shown here represents the largest component of the susceptibility of $\mathcal{H}_{\text{SI}}$ at $J_{\tilde\lambda}=0.4J_1$ where the system favors antiferromagnetic order.}
\label{fig:example_flow}
\end{figure}
The FRG equations are simultaneously solved on the imaginary frequency axis and in real space. A numerical solution requires (i) to discretize the frequency dependencies
and (ii) to limit the spatial dependence to a finite cluster, thus keeping correlations only up to some maximal length. In our calculations, the latter typically extends over distances of up to 9 lattice spacings corresponding to a correlation area (cluster size) of 181 lattice sites of the hexagonal lattice. The onset of spontaneous long-range order is signaled by a sudden breakdown of the smooth RG flow, while the existence of a stable solution indicates the absence of long-range order. (See Refs.~\onlinecite{reuther-10prb144410,reuther-11prb024402} for further technical details.) Fig.~\ref{fig:example_flow} shows an example for the characteristic flow behavior in a magnetically ordered phase.

\section{Results}\label{sec:results}

\subsection{Kane--Mele spin model}

From Eq.~(\ref{kanemelespinmodel}), the Kane--Mele spin model reduces to an isotropic nearest neighbor spin system in the limit of vanishing spin orbit coupling $J_{\lambda}=0$. In this case, the system exhibits the standard N\'eel state on the honeycomb lattice. Within our PFFRG approach, this type of order is signaled by an instability breakdown in the RG flow occurring at the $K$- and $K'$-points, i.e. the corners of the extended (second) Brillouin zone of the honeycomb lattice. (Unless stated otherwise, we plot the susceptibility in the second Brillouin zone of the underlying two-atomic Bravais lattice because the experimentally connected unfolded susceptibility has the periodicity of this extended zone.) Hence, at an RG scale right before the magnetic order sets in, the momentum resolved susceptibility shows pronounced peaks at the $K$- and $K'$-point positions. As a consequence of rotational invariance the susceptibility profile is identical for all directions of external magnetic fields.
\begin{figure}[t]
\centering
\includegraphics[scale=0.3]{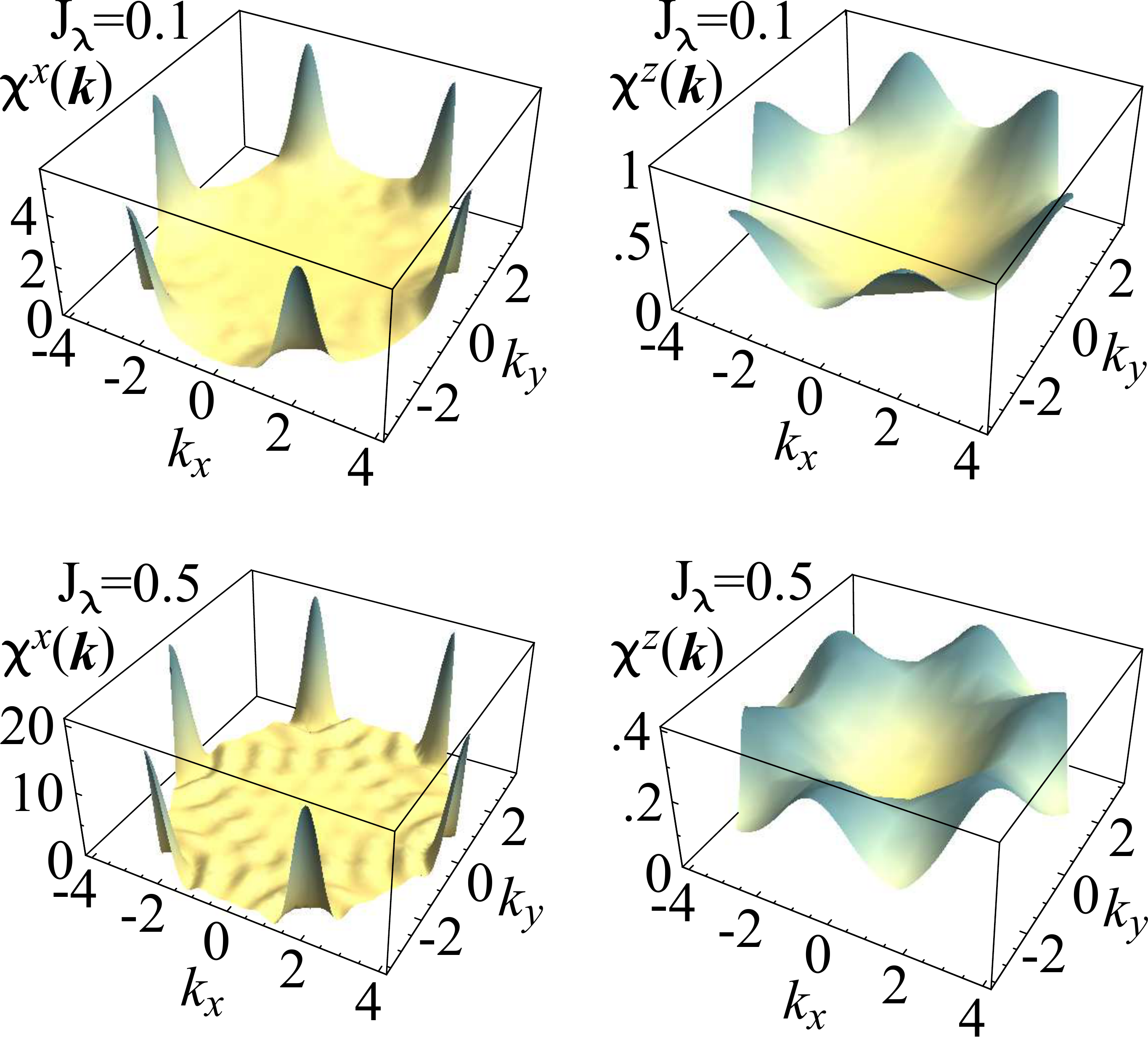}
\caption{Magnetic susceptibilities at the critical scale $\Lambda=\Lambda_{\text{c}}$ for various values of $J_{\lambda}$ ($J_1=1$) in the Kane--Mele spin model, resolved for in plane ($x,y$) and out of plane ($z$). Top row: $\chi^x(\bs{k})$ (left panel) and $\chi^z(\bs{k})$ (right panel) for $J_{\lambda}=0.1$. Bottom row: $\chi^x(\bs{k})=\chi^y(\bs{k})$ (left panel) and $\chi^z(\bs{k})$ (right panel) for $J_{\lambda}=0.5$. The susceptibility weight along $z$ significantly decreases for large $J_{\lambda}$. For higher $J_{\lambda}$, the remainder $z$-susceptibility deviates from the N\'eel AFM structure.}
\label{fig:km-suscep}
\end{figure}%
Once the spin orbit interaction $J_{\lambda}$ is switched on, the situation changes considerably as shown in Fig.~\ref{fig:km-suscep}. While the susceptibility peaks for an external field in $x$-direction (or $y$-direction) become even sharper as compared to $J_{\lambda}=0$, the peaks in the $z$-component drop drastically. Already at small $J_{\lambda}=0.1$ this effect is rather pronounced, which evidences that for finite $J_{\lambda}$, the spins favor the $x$-$y$ plane. (We set $J_1=1$ in this section.) With increasing $J_{\lambda}$, more weight of the $z$-susceptibility is transferred to the $x$- and $y$-components of $\chi$. For strong enough $J_{\lambda}$, the remnant magnetic fluctuations in $\chi^z$ are not of antiferromagnetic type anymore, which can be seen in Fig.~\ref{fig:km-suscep} for $J_{\lambda}=0.5$ showing small maxima at $M$-point positions.
We do not observe any particular phase transition at $\lambda>0$. In particular, the magnetic order persists in the whole parameter space. The frustration generated by the $J_{\lambda} S^z_i S^z_j$-terms has little effect because the spins can circumvent this frustration by avoiding the $z$-axes. With increasing $J_{\lambda}$, the two sublattices become effectively decoupled such that in the limit $J_{\lambda}\rightarrow\infty$ both sublattices exhibit $xy$ ferromagnetic order independently.

\subsection{Sodium iridate spin model}
\label{sec-si}
\begin{figure*}[t]
\centering
\includegraphics[scale=0.28]{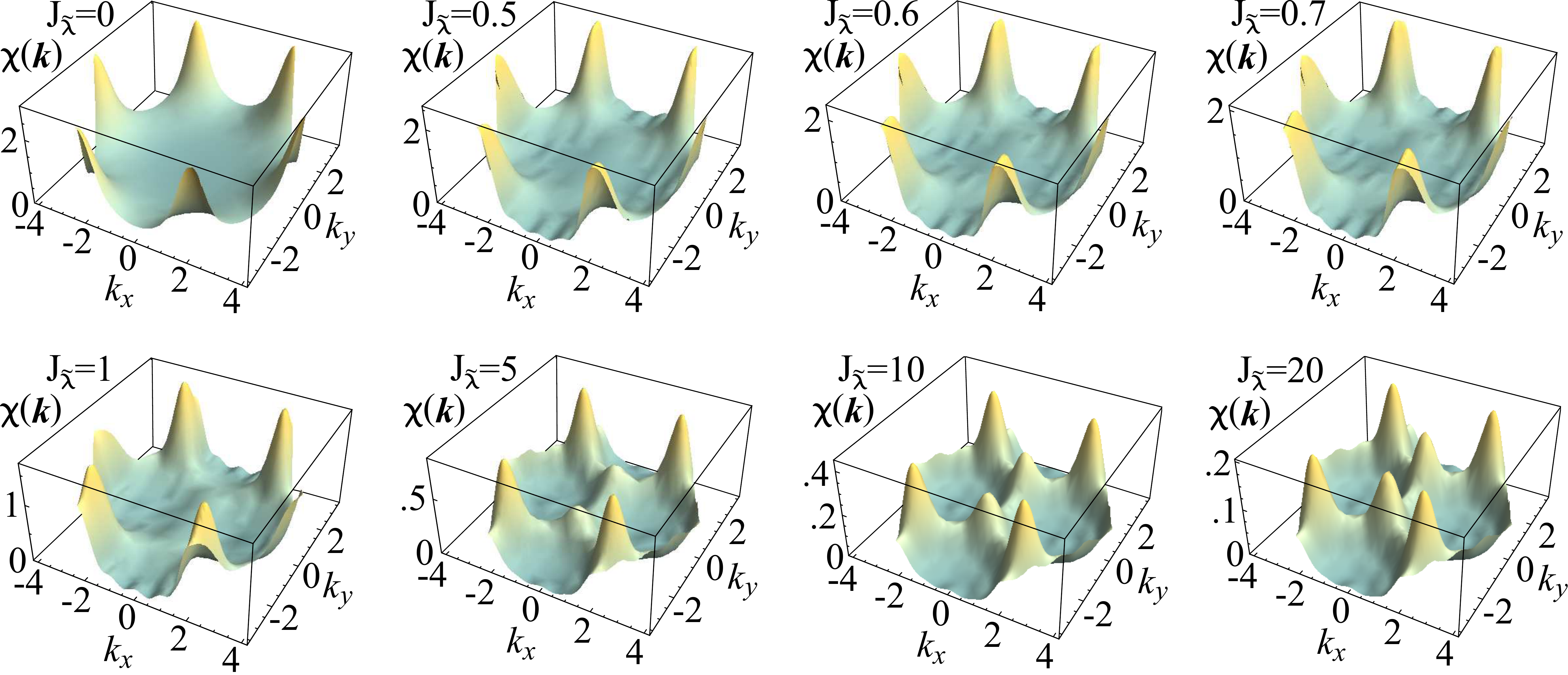}
\caption{Magnetic susceptibilities of the SI spin model for various values of $J_{\tilde\lambda}$ ($J_1=1$). All susceptibilities shown refer to a magnetic field in $z$-direction. The $x$- and $y$-components of the susceptibility are obtained by $k$-space rotations of $120^\circ$ in clockwise or counterclockwise direction, respectively. (See Section~\ref{sec-si} for more details.) While the N\'eel peaks initially persist for finite $J_{\tilde\lambda}$, the peaks start to move due to the onset of incommensurability (Fig.~\ref{fig:wave-vector}). For large $J_{\tilde\lambda}$, new suceptibility peaks emerge which link to the change of unit cell structure of magnetic order.}
\label{fig:shitade-suscep}
\end{figure*}
As for the KM spin model in the previous section, the SI spin model becomes a simple isotropic nearest neighbor spin system in the limit $J_{\tilde\lambda}=0$ and hence shows N\'eel order (upper left plot in Fig.~\ref{fig:shitade-suscep}). For finite but not too large $J_{\tilde\lambda}$, the antiferromagnetic order persists, i.e. the position of the ordering peaks in the susceptibility remains unchanged ($J_{\tilde\lambda}=0.5$ in Fig.~\ref{fig:shitade-suscep}). As the susceptibility looses its sixfold rotation symmetry for finite $J_{\tilde\lambda}$, this manifests in the deformation of the ordering peaks as compared to $J_{\tilde\lambda}=0$. Note that due to the special connection between lattice directions and spin directions in the SI spin model, the $x$-, $y$- and $z$-components of the susceptibility transform into each other under $k$-space rotations of $120^\circ$ in clockwise direction. Fig.~\ref{fig:shitade-suscep} illustrates $\chi^z$ which preserves the symmetries $k_x\rightarrow-k_x$ and $k_y\rightarrow-k_y$. Note that regardless of the particular phase, the value of the susceptibility at the six $K$- and $K'$-points must always be equal. This results from the fact that the three $K$-points (or $K'$-points) are related by reciprocal lattice vectors among each other. Furthermore, since the two sublattices are equivalent, the $K$- and $K'$-points are likewise degenerate. 

An interesting observation can be made regarding the orientation of the antiferromagnetic order. Due to the equivalence of the $x$-, $y$- and $z$-direction in spin space, the magnetic order can point in each of these directions without any preference. Even though SU(2) symmetry is explicitly broken, the rotational symmetry of the susceptibility prevails: consider a magnetic field ${\bf B}={\bf v}B$ pointing in some direction ${\bf v}=\sum_{\mu=x,y,z}v_{\mu}{\bf e}_{\mu}$ with $|{\bf v}|=1$. The corresponding susceptibility $\chi^{\bf v}$, i.e. the linear response to such a perturbation is defined as
\begin{align}
\chi^{\bf v}&=\frac{\partial {\bf Mv}}{\partial B}\Big{|}_{B\rightarrow0}=\frac{\partial(\sum_{\mu=x,y,z}M_{\mu}v_{\mu})}{\partial B}\Big{|}_{B\rightarrow0}\\ \notag
&=\sum_{\mu'=x,y,z}\frac{\partial(\sum_{\mu=x,y,z}M_{\mu}v_{\mu})}{\partial B_{\mu'}}\frac{\partial B_{\mu'}}{\partial B}\Big{|}_{B\rightarrow0}\\ \notag
&=\sum_{\mu,\mu'=x,y,z}v_{\mu}\chi^{\mu\mu'}v_{\mu'}\,,
\end{align}
where $\chi^{\mu\mu'}=\frac{\partial M_{\mu}}{\partial B_{\mu'}}\big{|}_{B\rightarrow0}$ and $\bf M$ is the magnetization. Since $\chi^{\mu\mu'}$ cannot develop any off-diagonal elements before reaching the magnetic instability in the RG flow,\cite{footnote} we have $\chi^{\mu\mu'}=\delta_{\mu\mu'}\chi^{\mu}$. It follows that
\begin{equation}
\chi^{\bf v}=\sum_{\mu=x,y,z}v_{\mu}^2\chi^{\mu}\,.
\end{equation}
Since $\chi^x=\chi^y=\chi^z$ at all $K^{(')}$-points, we obtain
\begin{equation}
\chi^{\bf v}_{K^{(')}} =\chi^{z}_{K^{(')}}\sum_{\mu=x,y,z}v_{\mu}^2=\chi^{z}_{K^{(')}}\,.
\end{equation}
Hence, in linear response the low energy physics of the system is rotationally symmetric and the antiferromagnetic order can point in any direction. This is a consequence of the N\'eel order residing at high-symmetry points of the Brillouin zone as well as the special connection between lattice directions and spin directions in the SI spin model. However, this argument does not hold for spin fluctuations away from the $K$- or $K'$-points. For fluctuations at arbitrary momentum, a certain direction will generally be preferred.

\begin{figure}[t]
\centering
\includegraphics[scale=0.66]{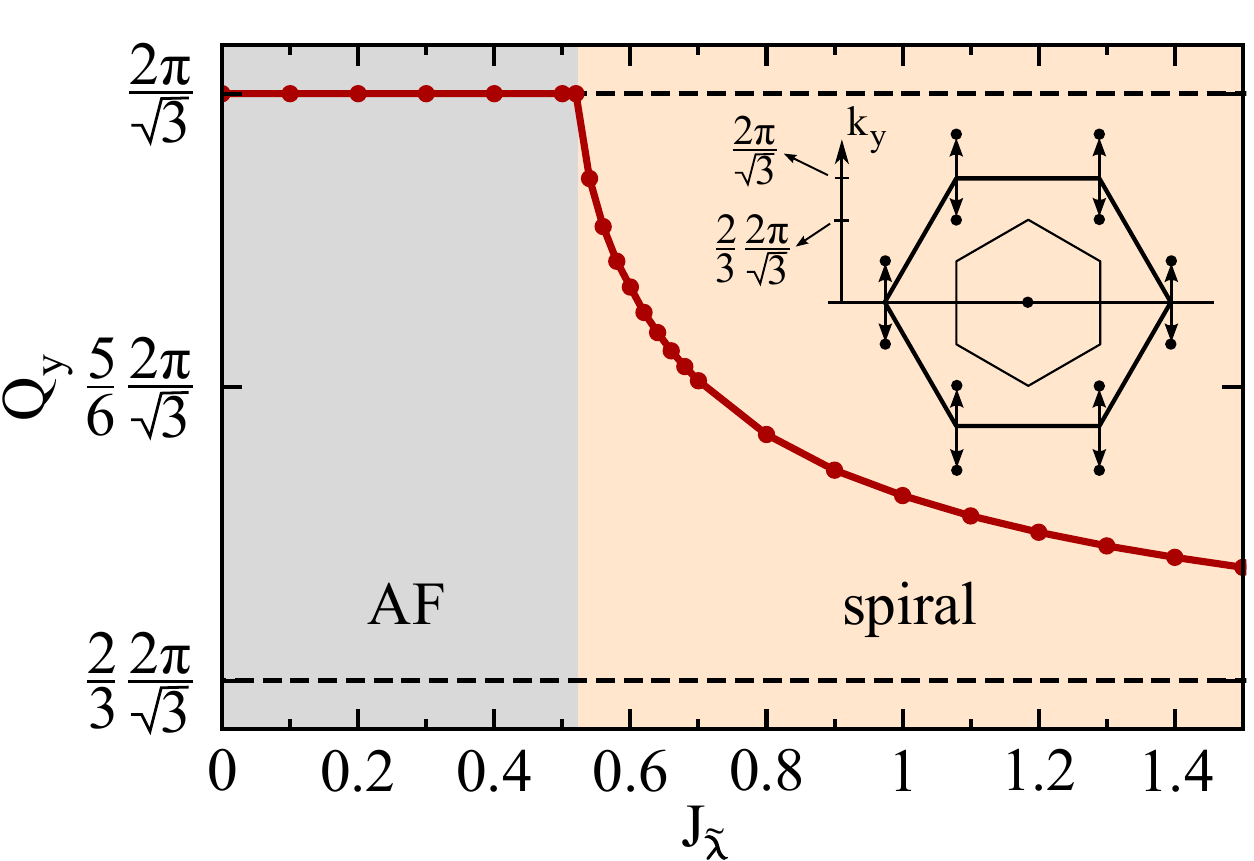}
\caption{Dependence of the ordering vector $Q_y$ on $J_{\tilde\lambda}$ in $\mathcal{H}_{\text{SI}}$. The inset illustrates the evolution of the ordering peaks in the Brillouin zone (thick hexagon: second Brillouin zone, thin hexagon: first Brillouin zone; see also Fig.~\ref{fig:shitade-suscep}). In the limit $J_{\tilde\lambda}\rightarrow \infty$, the system converges again towards a commensurate ordering vector.}
\label{fig:wave-vector}
\end{figure}

As $J_{\tilde\lambda}$ increases, the deformation of the ordering peaks at the $K$- and $K'$-points becomes more pronounced. At some coupling $J_{\tilde\lambda}\approx0.53$, the peaks split and the new maxima move along the $k_y$-direction (Fig.~\ref{fig:wave-vector}). 
These peak positions indicate a phase transition to a spiral phase with incommensurate order. It is important to note, however, that magnetic order persists in the whole parameter regime around the transition and we find no magnetically disordered phase. This can be seen from the behavior of the RG flow which always exhibits a characteristic instability breakdown.
To demonstrate the evolution of the ordering vector in the spiral phase, Fig.~\ref{fig:wave-vector} shows the peak position as function of $J_{\tilde\lambda}$. Note that the $k_x$-component of the peak position is constant in $J_{\tilde\lambda}$. With increasing $J_{\tilde\lambda}$, the peaks move continuously towards the points ${\bf Q}_{\infty}=(\pm\frac{2\pi}{3},\pm\frac{2}{3}\frac{2\pi}{\sqrt{3}})$ which lie at two third of the distance between the $K^{(')}$-points and the $k_x$-axis (Fig.~\ref{fig:wave-vector}). Again, with increasing $J_{\tilde\lambda}$ there is no sign of any non-magnetic phase.

\begin{figure}[t]
\centering
\includegraphics[scale=0.33]{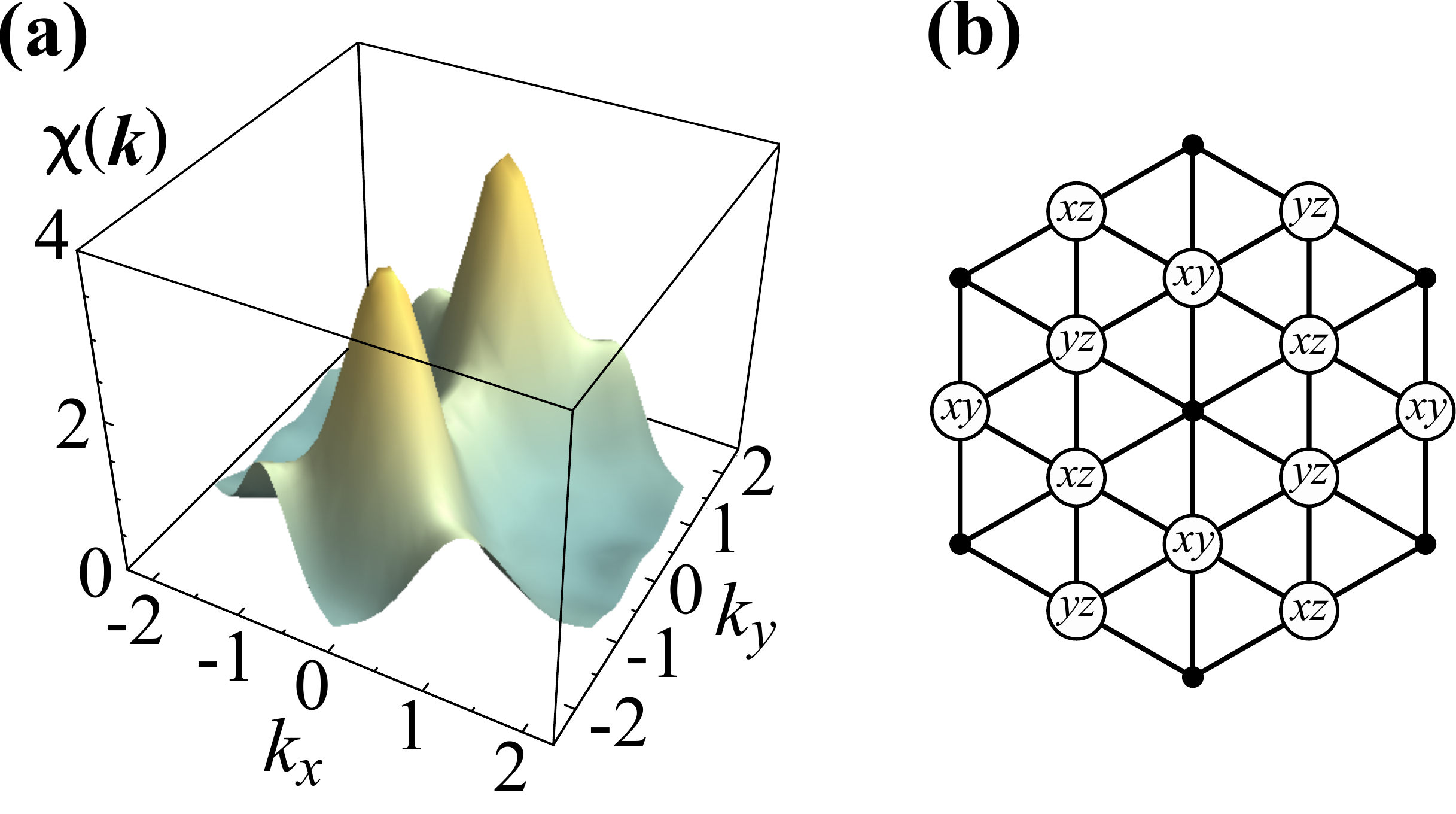}
\caption{The SI spin model at $J_{\tilde{\lambda}}\rightarrow\infty$: (a) Magnetic susceptibility displayed in the first Brillouin zone of the triangular sublattice. The two ordering peaks correspond to the peaks in Fig.~\ref{fig:shitade-suscep} which emerge at $J_{\tilde{\lambda}}\gtrsim5$ and $k_x=0$. In the limit $J_{\tilde{\lambda}}\rightarrow\infty$, these maxima reach the same hight as the ones at ${\bf Q}_{\infty}=(\pm\frac{2\pi}{3},\pm\frac{2}{3}\frac{2\pi}{\sqrt{3}})$. (b) Mapping of the SI spin model at $J_{\tilde\lambda}\rightarrow\infty$ to the antiferromagnetic Heisenberg model on the triangular lattice: The lattice is divided into four sublattices denoted by "$\bullet$", "$xy$", "$xz$" and "$yz$". As shown in Eq.~(\ref{transformation}) the transformation from ${\bf S}_i$ to $\tilde{\bf S}_i$ depends on the sublattice where $i$ resides. The exchange couplings follow the convention shown in Fig.~\ref{fig:soc-hoppings}.}
\label{fig:so_infty}
\end{figure}

The system at infinite spin-orbit coupling is of particular interest, as this case represents a model with Kitaev-like interactions on the triangular lattice. As $J_{\tilde\lambda}$ goes to infinity, the system is effectively described by decoupled triangular sublattices. Hence, already the first Brillouin zone, i.e. the Brillouin zone of a triangular sublattice, contains the full information about $\chi$ in $k$-space. The susceptibility then becomes periodic with respect to this smaller zone. Such a change of periodicity can be seen in Fig.~\ref{fig:shitade-suscep} at large $J_{\tilde\lambda}$ where new peaks at $k_x=0$ emerge. In the limit $J_{\tilde\lambda}\rightarrow\infty$ these new peaks reach the same height as the ones at ${\bf Q}_{\infty}$ and finally become {\it identical} to them, indicating the new periodicity in $k$-space. Fig.~\ref{fig:so_infty}a shows the susceptibility in the first Brillouin zone of the triangular sublattice in this limit. From the peak positions, one can easily derive the corresponding spin pattern in real space. On each triangular sublattice the wave vector is half the one of the $120^{\circ}$-N\'eel order residing at the corners of the first Brillouin zone. The unit cell contains $6\times6$ lattice sites as compared to the $3\times3$ unit cell of the $120^{\circ}$-N\'eel order. Hence, the order is commensurate and the local magnetic moments along a lattice direction are modulated with a periodicity of 6 sites. Taking into account both sublattices of the honeycomb lattice, we end up with a unit cell containing 72 sites. 

Our numerical conclusions for the SI spin model in the limit $J_{\tilde\lambda}\rightarrow\infty$ can also be reconciled with an analytical argument. Performing a transformation in spin space, ${\bf S}_i\rightarrow \tilde{\bf S}_i$, the system at this point can be mapped to an SU(2) invariant antiferromagnetic Heisenberg model on the triangular lattice, $\mathcal{H}_{\text{SI}}=\sum_{ij}\tilde{\bf S}_i\tilde{\bf S}_j$. For this mapping, we divide the triangular lattice into four sublattices denoted by "$\bullet$", "$xy$", "$xz$" and "$yz$", each with a doubled lattice constant (Fig.~\ref{fig:so_infty}b). The relation between ${\bf S}_i$ and $\tilde{\bf S}_i$ depends on the sublattice,
\begin{align}
i\in"\bullet":\quad&\tilde{\bf S}_i=(S_i^x,S_i^y,S_i^z)\,,\notag\\
i\in"xy":\quad&\tilde{\bf S}_i=(-S_i^x,-S_i^y,S_i^z)\,,\notag\\
i\in"xz":\quad&\tilde{\bf S}_i=(-S_i^x,S_i^y,-S_i^z)\,,\notag\\
i\in"yz":\quad&\tilde{\bf S}_i=(S_i^x,-S_i^y,-S_i^z)\,,\label{transformation}
\end{align}
e.i., while on sublattice "$\bullet$" the spins remain unchanged, on the sublattice "$xy$" the $x$- and $y$-components of the spin operator acquire a minus sign, and so on (a similar mapping for the Heisenberg-Kitaev model at $\alpha=0.5$ is described in Ref.~\onlinecite{chaloupka-10prl027204}). Since the antiferromagnetic Heisenberg model on the triangular lattice exhibits magnetic order via the $120^{\circ}$-N\'eel state~\cite{capriotti-prl3899}, it follows that the SI spin model at $J_{\tilde\lambda}\rightarrow\infty$ is likewise magnetically ordered. The corresponding spin pattern in real space can be found by applying the inverse of the above spin transformation to the $120^{\circ}$-N\'eel state: Since the structure of the spin rotations (Fig.~\ref{fig:so_infty}b) has a periodicity of two lattice sites in each lattice direction, the $3\times3$ unit cell of the $120^{\circ}$-N\'eel order transforms back into a $6\times6$ unit cell, as found within our PFFRG calculations.

\section{Discussion}\label{sec:discussion}
\begin{figure}[t]
\centering
\includegraphics[scale=0.75]{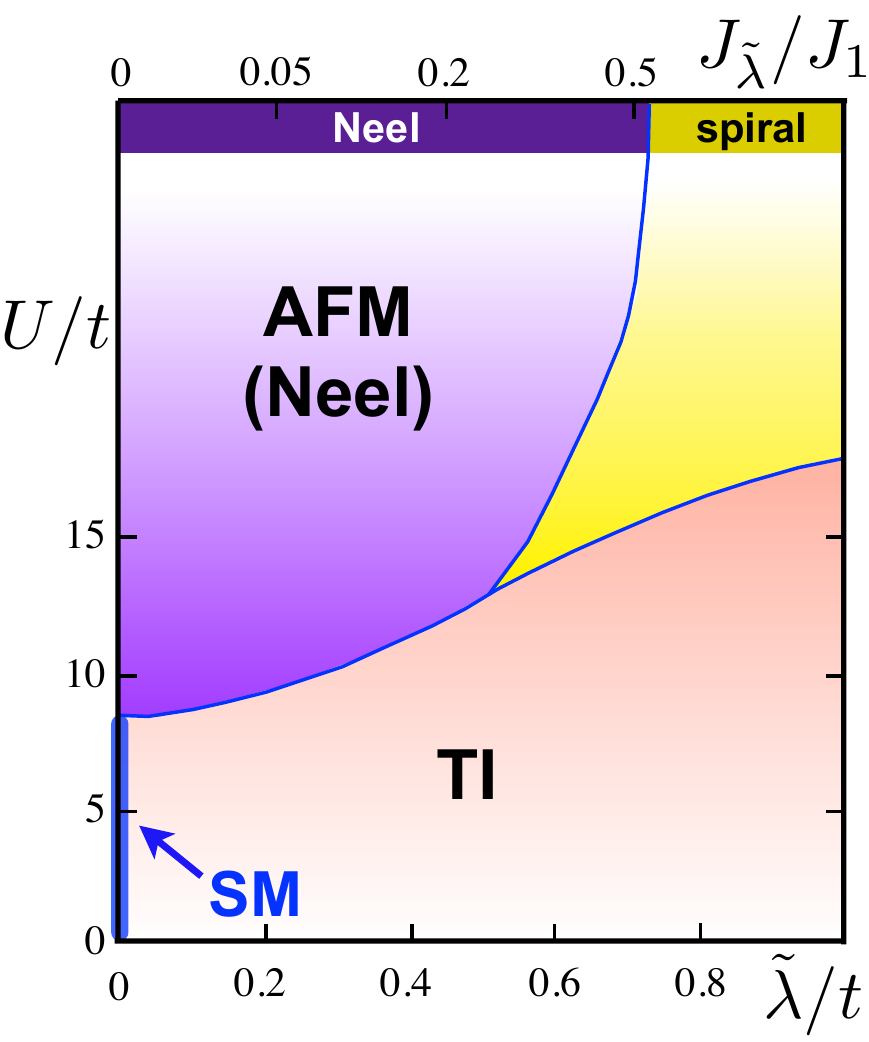}
\caption{(color online) Schematic phase diagram for the SI Hubbard model as conjectured from our strong coupling results of a N\'eel to spiral transition at $J_{\tilde\lambda}= 0.53$ (upper $x$ scale), corresponding to $\tilde\lambda/t \approx 0.73$ (lower $x$ scale) . While we cannot ultimately assess the nature of a possible intermediate phase aside from TI and AFM for strong interaction and strong spin-orbit coupling (yellow phase), its existence is likely.}
\label{fig:res-pd}
\end{figure}
In view of our results for the SI model, we speculate about the implications for the phase diagram at intermediate $U$. We find the incommensurate phase for $J_{\tilde\lambda}/J_1\geq 0.53$ where $J_{\tilde\lambda}=4{\tilde\lambda}^2/U$ and $J_1=4t^2/U$, implying a transition at $\frac{\tilde\lambda}{t}\approx 0.73$ for large $U$.
In Fig.~\ref{fig:res-pd}, we have replotted the phase diagram of R\"uegg and Fiete\cite{ruegg-12prl046401} in a slightly modified way. Our reasoning is the following: since the spin model corresponds to $U\rightarrow \infty$, we extrapolated the phase boundary between ``VBS (AFM)''  and ``QSH*'' of the phase diagram in Ref.~\onlinecite{ruegg-12prl046401} to larger $U$. As the phase transition occurs for sufficiently large $U$ approximately at $\frac{\tilde\lambda}{t}\approx0.73$ (Fig.\,\ref{fig:res-pd}), we speculate that the observed phase transition from N\'eel to spiral order in the spin model is a remnant of the phase transition into the QSH* phase at intermediate $U$. Within this scenario, the QSH* phase would transform into spiral magnetic order in the limit $U\rightarrow\infty$. We note, however, that this necessarily implies that with increasing $U$ the gap of the QSH* phase closes at some point to form the Goldstone mode of the spiral order. In principle, the gap closure can occur at finite $U$ (which would imply an additional phase boundary in Fig.~\ref{fig:res-pd}) or at $U\rightarrow \infty$. In the latter scenario, the QSH* phase could extend up to $U\rightarrow\infty$. Furthermore, one should keep in mind that the QSH* phase as found in Ref.~\onlinecite{ruegg-12prl046401} might not be the only topological liquid candidate with similar properties, such as a doubled semion spin liquid~\cite{scharfenberger-11prb140404}. The alternative scenario---assuming that a QSH$^\star$-type phase does {\it not} exist---would still require an additional phase comparable to the yellow phase of the schematic phase diagram in Fig.\,\ref{fig:res-pd}); in this case, the additional phase would most likely be a magnetically ordered phase (\eg spiral order). Whether or not this phase is a topological liquid or just another magnetically ordered phase, we conjecture that in either case an additional phase of some kind should be present. 

In summary, we find that the physics of the SI spin model is much richer as compared to the KM spin model. This can be traced back to the different spin symmetries in both systems. The broken axial symmetry in the SI spin model prevents the spins from forming planar antiferromagnetic order and eventually leads to the emergence of a spiral phase. Note that this phase does not have any analogue in similar models such as the Heisenberg-Kitaev model.\cite{reuther-11prb100406}
As such, we have identified multi-directional spin orbit terms to be an interesting way to create new spin phases in the infinite $U$ limit and possibly even more exotic phases at intermediate $U$ when charge fluctuations enter the picture.

To give another direction of further investigation, it will be interesting to study the KM model in the presence of Rashba spin orbit coupling
\begin{equation*} 
H_{\text{R}}=i\lambda_R \sum_{\langle i j \rangle}\sum_{\alpha\beta} c_{i\alpha}^\dag [\hat e_z (\bs{\sigma}\times\bs{d}_{ij})]_{\alpha\beta} c_{j\beta}^\pd\ 
\end{equation*}
The Rashba term breaks the remaining U(1) symmetry of the electron spin to $\mathbb{Z}_2$, and also affects the $z\to -z$ mirror symmetry as well as particle--hole symmetry. 
Taking into account the Rashba spin orbit coupling results in a more complicated spin Hamiltonian with some terms being of Dzyaloshinskii-Moriya type. The full Hamiltonian is given by
\begin{widetext}
\begin{eqnarray}\label{ham:kmr-spin}
\mathcal{H}_{\rm KMR} &=&
J_\lambda \sum_{\ll ij \gg} \left[ -S_i^x S_j^x - S_i^y S_j^y + S_i^z S_j^z \right]\nonumber \\
&&+ \sum_{\delta_1-{\rm links}} \left[ ( J_1 + J_{\rm R} )S_i^x S_j^x 
+ (J_1 - J_{\rm R})(S_i^y S_j^y + S_i^z S_j^z) - \sqrt{ J_1 J_{\rm R}} ( S_i^y S_j^z  - S_i^zS_j^y ) \right]
\nonumber \\
&&+ \sum_{\delta_2-{\rm links}} \left[ S_{i/j}^x \to -\frac{1}{2} S_{i/j}^x - \frac{\sqrt{3}}{2} S_{i/j}^y~~\hbox{and}~~S_{i/j}^y \to \frac{\sqrt{3}}{2} S_{i/j}^x - \frac{1}{2} S_{i/j}^y  \right]
\nonumber \\
&&+ \sum_{\delta_3-{\rm links}} \left[ S_{i/j}^x \to -\frac{1}{2} S_{i/j}^x + \frac{\sqrt{3}}{2} S_{i/j}^y~~\hbox{and}~~S_{i/j}^y \to -\frac{\sqrt{3}}{2} S_{i/j}^x - \frac{1}{2} S_{i/j}^y  \right],
\end{eqnarray}
\end{widetext}
where the third line in~\eqref{ham:kmr-spin} is obtained from the second one by replacing $S_{i/j}^x$ by $-1/2 S_{i/j}^x - \sqrt{3}/2 S_{i/j}^y$ and so on. The different links denoted by $\delta_i$ $(i=1,2,3)$ are the three nearest neighor vectors of the honeycomb lattice. We expect the $J_{\lambda}$--$J_{\rm R}$ phase diagram to be interesting and to host some additional phases, which we defer to a future publication.

\section{Conclusion}\label{sec:conclusion}
We have investigated the strong coupling limit of Hubbard models of topological honeycomb band structures. We have considered two band structures both classified as two--dimensional $\mathbb{Z}_2$ topological insulators, where only the Kane--Mele spin model as opposed to the sodium iridate model preserves axial spin symmetry. For the former model at infinite coupling, the magnetism tends to form $XY$ antiferromagnetic order already at very small spin orbit couplings. This way the spins manage to avoid the frustration induced by the spin-orbit anisotropic spin terms. As a consequence, frustration effectively plays no role in the KM spin model. The physical scenario is very different for the sodium iridate model with generalized spin orbit couplings and hence broken axial spin symmetry. There, the spins cannot form $XY$, $XZ$ or $YZ$ order. As a result, the magnetic phase formation in the strong coupling limit exhibits a commensurate to incommensurate N\'eel to spiral transition at $J_{\tilde{\lambda}}\approx 0.53$. In the limit of infinite spin orbit coupling, the model converges to a commensurate magnetic state with a $6\times6$-site unit cell on each of the two decoupled triangular sublattices of the underlying honeycomb model. The emergence of the spiral phase in the infinite $U$ limit leads us to conjecture that aside from the topological band insulator regime and the antiferromagnetic phase, a third phase should exist at finite $U$ and finite spin orbit coupling. In this respect, our results are not inconsistent with the existence of a fractionalized QSH* phase as proposed in Ref.~\onlinecite{ruegg-12prl046401}. Whatever this phase will eventually turn out to be, we find that the breaking of axial spin symmetry is generally vital to the emergence of new phases and an enriched diversity of magnetic phases in interacting topological honeycomb band structures.   

{\it Note added.} Recently, the classical Kitaev-Heisenberg model has been studied on the triangular lattice within Monte Carlo approaches. Interestingly, spiral-like order has been found to be associated with a $Z_2$ vortex lattice.\,\cite{daghofer}

\begin{acknowledgments}
We acknowledge useful discussions with Karyn Le Hur, Gregory Fiete, Giniyat Khaliullin, George Jackeli, Andreas R\"uegg, and Matthias Vojta. JR is supported by the Deutsche Akademie der Naturforscher Leopoldina through grant LPDS 2011-14. RT is supported by an SITP fellowship by Stanford University. SR acknowledges support from DFG under Grant No.\ RA~1949/1-1. 
\end{acknowledgments}

\bibliography{strong-topo-prb}

\end{document}